\documentclass{article}
\usepackage{graphicx} 
\usepackage{amsmath}
\usepackage{hyperref}
\usepackage{authblk}
\begin{document}
\title{Comparison of pair-source and correlation-function based Lévy analyses in Ar+Sc system at SPS energies}

\author[1,2]{Barnabás Pórfy}
\affil[1]{Department of Atomic Physics, ELTE E\"otv\"os Lor\'and University, Budapest, Hungary}
\affil[2]{HUN-REN Wigner Research Centre for Physics, Budapest, Hungary}
\maketitle

\begin{abstract}
Non-Gaussian features in femtoscopic observables have been observed in high-energy collisions, yet their origin remains not fully understood. This study focuses on the phenomenological approach to femtoscopy, going beyond the Gaussian approximation by utilizing Lévy-stable source distributions. Using UrQMD simulations, three-dimensional analyses of pion sources are done in central Ar+Sc collisions, and the extracted parameters are studied as functions of transverse mass and collision energy. This analysis contributes to a hadronic baseline motivated by past and future research work in femtoscopy at the NA61/SHINE experiment. A direct methodological comparison is presented between parameters extracted from the projections of three-dimensional pair-source distribution and those obtained from the three-dimensional correlation function, reconstructed via Fast Fourier Transform.
\end{abstract}

\section{Introduction}
The main goal of high-energy heavy-ion experiments of today is exploring and understanding the properties of the quark--gluon plasma (QGP). One possible measurement is femtoscopy, the analysis of quantum-statistical correlations in heavy-ion collisions~\cite{Goldhaber:1959mj, Goldhaber:1960sf}, which uncovers the femtometer scale space-time structure of the particle-emitting source~\cite{Lisa:2005dd, Csorgo:1999sj}.
Recent experimental results in Refs.~\cite{PHENIX:2017ino,NA61SHINE:2023qzr,CMS:2023xyd,Kincses:2024sin,PHENIX:2024vjp} using Lévy shaped source~\cite{Csorgo:2003uv,Csanad:2007fr}, show that the previously assumed Gaussian shape of the source~\cite{Csorgo:1994fg, Akkelin:1995gh} does not describe the data well. The appearance of Lévy distributions in high-energy heavy-ion physics can be attributed to several physical mechanisms, such as jet fragmentation~\cite{Csorgo:2004sr}, critical phenomena~\cite{Csorgo:2005it}, however, the most prominent of them is the Lévy walk~\cite{Kincses:2024lnv}. These mechanisms all result in Lévy-distributed sources~\cite{Csorgo:2003uv,Csanad:2007fr}.

Experimentally, one cannot access the source distribution $S(r)$ directly ($r$ is the four-momentum distance vector); however, by utilizing femtoscopy, one can get information about the distribution via measurements of momentum correlations. Simulation-wise, event generators provide a great tool, as one can get the freeze-out coordinates of the last point of interaction of each produced particle. This enables the calculation of the freeze-out distance distributions:
\begin{equation}\label{e:pairsource}
	D(r,K) = \int d^4\sigma S(\sigma + r/2,K)S(\sigma - r/2,K),
\end{equation}
where $\sigma = (x_1 + x_2)/2$ is the pair center of mass, and $K = (p_1 + p_2)/2$ is the average pair momentum. In experimental analysis, the research focuses on the quantum-statistical correlation functions. The connection between the two-particle momentum correlations ($C_2$) and the pair source distribution ($D$) is as follows:
\begin{equation}\label{e:corrfunc}
	C_2(q,K) = \int d^4r D(r,K)|\Psi^{(2)}_q(r)|^2,
\end{equation}
where $q = p_1 - p_2$ is the momentum difference of the pair. Both Eq.~(\ref{e:corrfunc} and~\ref{e:pairsource}) depend on $K$, however, it is assumed to be weaker~\cite{Lisa:2005dd} than $q$ and often appears only through the dependence of parameters on it. As discussed already, the Lévy distributions provide a more general approach in describing the source shape, the three-dimensional

\begin{equation}
	\mathcal{L}(\alpha, R^2; \vec{r}) = \frac{1}{(2\pi)^3}
\int d^3\vec{\zeta} \, e^{i \vec{\zeta} \cdot \vec{r}} \, e^{-\frac{1}{2} \left| \vec{\zeta}^T R^2 \vec{\zeta} \right|^{\alpha/2}} = D(\vec{r}),
\label{e:Levydef}
\end{equation}
where $\alpha$ is the Lévy stability index, and $R^2$ is the matrix of Lévy scale parameters, often assumed to be diagonal, i.e., $\mathrm{diag}\left(R_{\mathrm{out}}^2, R_{\mathrm{side}}^2, R_{\mathrm{long}}^2\right)$ in the Bertsch--Pratt coordinate system~\cite{Bertsch:1988db, Pratt:1990zq}. Furthermore, $\vec r$ is the vector of spatial coordinates, and the vector $\vec \zeta$ represents the integration variable. For $\alpha=2$, the distribution is Gaussian, whereas $\alpha=1$ gives back the Cauchy distribution; for $\alpha < 2$, the Lévy distribution exhibits a power-law tail. If single-particle sources are Lévy distributed, the pair source distribution retains the same source shape parameters under convolution~\cite{PHENIX:2017ino}.
Finally, the correlation function introduced in Eq.~\ref{e:corrfunc} has the following shape without final state interactions:
\begin{equation}
C_2(q) = \int d^4r D(r,K)|\Psi^{(2)}_q(r)|^2 = 1 + \lambda\cdot e^{-|q^TR^2q|^{\alpha/2}}.
\label{e:fittingformulanocoul}
\end{equation}
This final formula shows that $C(q)$ can be constructed from the pair-source distribution with Fourier transformation, e.g., the Fast Fourier Transform (FFT).

In this paper, the investigation of the two-particle source functions and correlation functions constructed using the FFT is presented. The analysis uses like-charged pions in central $^{40}$Ar+$^{45}$Sc collisions, one of the collision systems used in NA61/SHINE experiment~\cite{Abgrall:2014fa}, at the available energy range ($\sqrt{s_\text{NN}} \approx $5-17 GeV, or 13-150A GeV/$c$ beam momentum) generated by the UrQMD framework~\cite{Bass:1998ca, Bleicher:1999xi}. The reconstructed two-particle source functions are then fitted with three-dimensional Lévy distributions~\cite{LevyCode}, similarly to Refs.~\cite{Kincses:2024lnv, Kincses:2025iaf, Kincses:2025izu, Huang:2025edi, Porfy:2025sev, Molnar:2026vgi, Kincses:2026tfa}. For the correlation function, all available events were used (the same events as for $D(\rho)$) and then fitted with the three-dimensional Lévy-shaped correlation function~\cite{Csorgo:2003uv}. The source parameters are studied as functions of pair transverse mass ($m_\text{T}$) and collision energy.

\section{Analysis Details}
Using the UrQMD microscopic transport model, data was generated for the $^{40}$Ar+$^{45}$Sc system, motivated by the NA61/SHINE experiment~\cite{Abgrall:2014fa}. A total of 10,000 three-dimensional events were generated for the collision system, at beam momenta of 13$A$, 19$A$, 30$A$, 40$A$, 75$A$, and 150$A$ GeV/$c$, with a 0--10\% centrality cut based on forward energy, detailed in Ref.~\cite{Porfy:2025sev}. The generator was used in cascade mode, slightly altered to include the decay of $\eta$ mesons, with modifications based on~\cite{Uzhinsky:2013ata}, similarly to Ref.~\cite{Porfy:2025sev}. In the following subsections, two methods for extracting Lévy parameters from simulated events are presented. First, the pair-source distribution is briefly discussed, an experimentally inaccessible quantity; then the reconstruction of the momentum correlation function is detailed, providing the direct link to observable experimental data. 

\subsection{Pair source function}
To access the pair source function, the longitudinally comoving system (LCMS) alongside the Bertsch--Pratt (out-side-long) coordinate system~\cite{Pratt:1986ev,Bertsch:1988db} was selected. The cross-terms of $R^2$ in Eq.~\eqref{e:Levydef} are assumed to vanish at mid-rapidity~\cite{Chapman:1994yv, PHENIX:2004yan}, rendering the source nearly spherically symmetric~\cite{PHENIX:2017ino}. The three-dimensional spatial distance vector, $\vec{\rho} = \vec{r} - (\vec{K}/K_0) t$, boosted to the LCMS and its components are:
\begin{align}
	\rho_\text{out}^\text{LCMS} &= r_x \cos \varphi + r_y \sin \varphi - \frac{K_\text{T}}{K^2_0 - K_z^2} \left(K_0t - K_z r_z\right) \\
	\rho_\text{side}^\text{LCMS} &= -r_x \sin \varphi + r_y \cos \varphi \\
	\rho_\text{long}^\text{LCMS} &= \frac{K_0 r_z - K_z t}{\sqrt{K^2_0 - K_z^2}},
\end{align}
where $\cos \varphi = K_x/K_\text{T}, \sin \varphi = K_y/K_\text{T}$, and $K_\text{T}^2 = K_x^2 + K_y^2$. This approach is the same as the one used in Refs.~\cite{Kincses:2024lnv, Kincses:2025iaf, Kincses:2025izu, Huang:2025edi, Porfy:2025sev, Molnar:2026vgi, Kincses:2026tfa}. The one-dimensional projections using $\rho$ spatial distance vectors are $D(\rho_\text{out}^\text{LCMS})$, $D(\rho_\text{side}^\text{LCMS})$, and \\$D(\rho_\text{long}^\text{LCMS})$. These are then calculated and fitted with the 3D Lévy distribution defined in Eq.~\eqref{e:Levydef}.

Because single events in the measured systems lack sufficient statistics for stable spatial fits~\cite{Korodi:2022ohn,Kincses:2022eqq}, the $D(\vec{\rho})$ distributions are constructed by summing over event blocks~\cite{Kincses:2025izu, Huang:2025edi, Porfy:2025sev, Molnar:2026vgi, Kincses:2026tfa}, averaging the results across blocks. To reproduce experimental conditions~\cite{Porfy:2024jsu}, kinematic and momentum difference cuts are applied. The specific number of merged events ($N_{\rm events}$) and the exact kinematic cuts applied are summarized in Table~\ref{tab:kinematic_cuts}, taken from Ref.~\cite{Porfy:2025sev}.
\begin{table}[]
 \scriptsize
 \caption{Summary of event merging, kinematic cuts, and $K_\text{T}$ binning.}
 \label{tab:kinematic_cuts}
 \centering
 \begin{tabular}{c|c|c|c|c|c}
 \hline
 System & $N_{\text events}$ blocks & Rapidity [$y$] & $p_\text{T}$ & $q_\text{LCMS}$ [GeV/$c$] & $K_\text{T}$ [GeV/$c$] \\ \hline\hline
 $^{40}$Ar+$^{45}$Sc & \begin{tabular}[c]{@{}c@{}}$p_\text{beam} \ge 40A$: 2500\\ $p_\text{beam} < 40A$: 5000\end{tabular} & $\pm 1$ ($\pm 1.5$ at $150A$) & $< 1.5$ & $< \sqrt{m_\text{T} \cdot 150 \text{ MeV}/c^2}$ & 8 bins $\in [0, 1]$ \\ \hline
 \end{tabular}
\end{table}

\subsection{Correlation function}
As discussed earlier, this work presents simulations and subsequent measurements of correlation functions in the $^{40}$Ar+$^{45}$Sc system. While experiments directly measure the correlation function, they have no access to the pair-source distribution. Another approach to compare simulated and experimental results is to reconstruct the correlation function, i.e., perform Eq.~\eqref{e:corrfunc}. For this approach, first, the three-dimensional pair-source distribution is constructed, $D(\rho)$, using all the simulated data. Then $D(\rho)$ is transformed into momentum space to create the three-dimensional correlation function. To construct $D(\rho)$, the same centrality selections, kinematic cuts, and $K_\text{T}$ binning were used as in the experimental data; these selections are then present in the correlation function $C(q_\text{out}, q_\text{side}, q_\text{long})$ as well. To create $D(\rho)$, an odd number of bins is used in all three spatial directions, allowing the central $\rho$ bin to be positioned exactly at zero. Notably, this method utilizes the full 3D correlation function, rather than the 1D projections typically employed in $D(\rho)$ analyses.

\begin{figure}[]
\centerline{\includegraphics[width=1.0\textwidth]{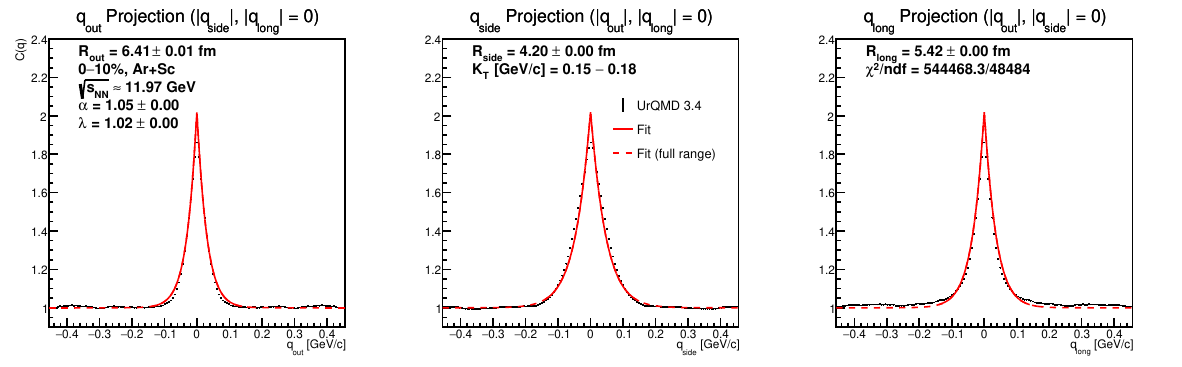}}
\caption{One-dimensional projections (``out'', ``side'', ``long'') of the three-dimensional correlation function and fit for $\pi^\pm\pi^\pm$ pairs in 0--10\% central Ar+Sc collisions at $p_\text{beam} = 75A$ GeV/$c$.}
\label{fig:exfit}
\end{figure}

Upon obtaining the three-dimensional correlation functions for each $K_\text{T}$, they were fitted with Eq.~\eqref{e:fittingformulanocoul}, where the squared radii matrix $R^2$ is assumed to be diagonal. The five fit parameters ($\alpha$, $\lambda$, $R_\text{out}$, $R_\text{side}$, and $R_\text{long}$) and their associated uncertainties were estimated using the Minuit2 and MINOS packages with the ROOT framework~\cite{ROOT_NIMA_1997}. Furthermore, the fits were restricted to a spherical momentum range defined by $|\vec{q}| < q_\text{max}$, where $|\vec{q}| = \sqrt{q_\text{out}^2 + q_\text{side}^2 + q_\text{long}^2}$. The specific $q_\text{max}$ values for each $K_\text{T}$ bin are provided in Table~\ref{tab:ArScfit_ranges_qmax}. An example of a 1D projection of the 3D fit is shown in Fig.~\ref{fig:exfit}.

\begin{table}[h!]
\centering
\footnotesize
\caption{The maximal values of $q$ defining the spherical fit range ($|\vec{q}| < q_\text{max}$). These values are defined separately for each $K_\text{T}$ bin but remain constant across the different beam momenta.}
\label{tab:ArScfit_ranges_qmax}
\begin{tabular}{cccccccc}
\hline
\multicolumn{8}{c}{$q_\text{max}$ (GeV$/c$) in given $K_\text{T}$ (GeV/$c$) bin} \\
$0.00$--$0.07$ & $0.07$--$0.11$ & $0.11$--$0.15$ & $0.15$--$0.18$ & $0.18$--$0.23$ & $0.23$--$0.30$ & $0.30$--$0.50$ & $0.50$--$1.00$ \\ \hline\hline
0.14 & 0.12 & 0.12 & 0.14 & 0.16 & 0.16 & 0.18 & 0.18 \\
\hline
\end{tabular}
\end{table}

\section{Results}
The results presented in this section are based on simulations generated by the UrQMD 3.4 model at SPS energies that match those measured by the NA61/SHINE experiment. A direct methodological comparison is presented for the Ar+Sc system, showing parameters obtained from the pair-source distributions and the correlation functions as functions of transverse mass, $m_\text{T} = \sqrt{K_\text{T}^2 + m^2}$, utilizing the same set of simulated events, but instead of block averaging, an average over all events is done. It is important to highlight that the correlation function results do not yet include a systematic uncertainty study.

\begin{figure}[]
\centerline{\includegraphics[width=1.0\textwidth]{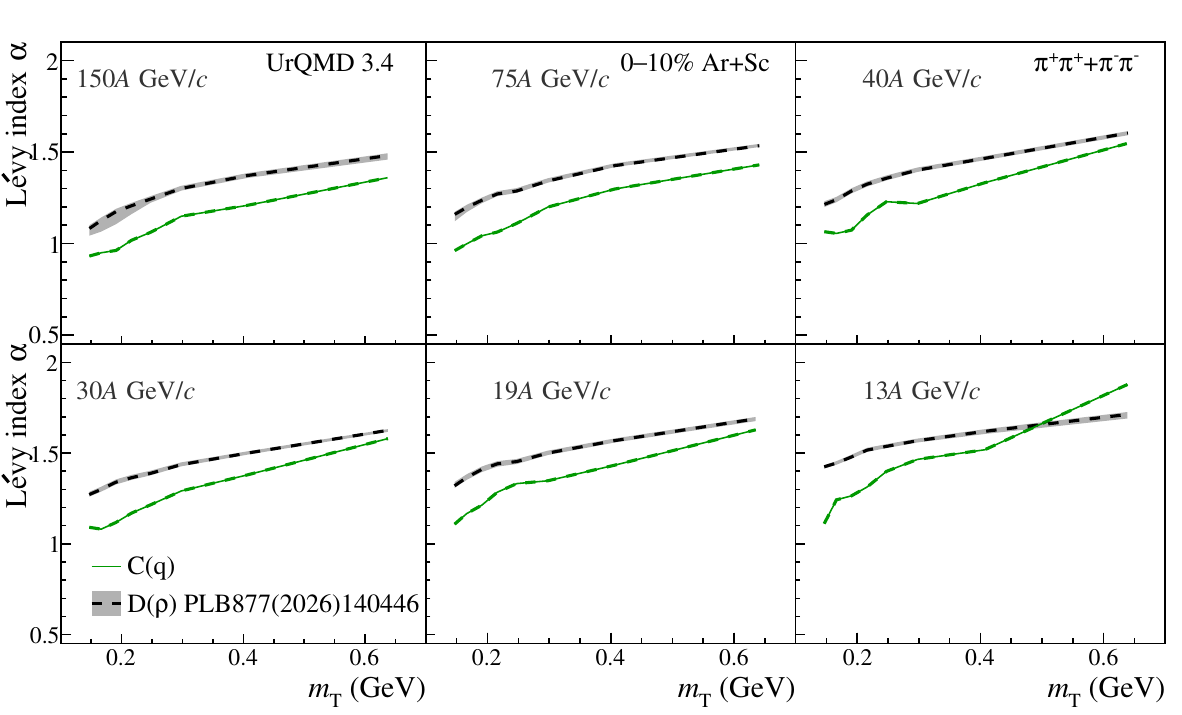}}
\caption{The Lévy index $\alpha$ for 0--10\% central collisions in Ar+Sc as a function of $m_\text{T}$. The dashed lines represent values extracted from the simulated $D(\rho)$ distributions, while the solid lines represent the $C(q)$ results. The colored bands denote systematic uncertainties.}
\label{fig:fftvsDrhoA}
\end{figure}

The Lévy index $\alpha$, shown in Fig.~\ref{fig:fftvsDrhoA}, exhibits a mild dependence on beam momentum, similar to the $D(\rho)$ results presented in Ref.~\cite{Porfy:2025sev}; however, the values extracted from $C(q)$ are systematically smaller than those from $D(\rho)$. Towards higher $m_\text{T}$ values, the two methods converge, and in the case of $13A$ GeV/$c$, the parameter from the correlation function method surpasses that of the $D(\rho)$ method. This mild decrease is observed in both simulation methods and is in contrast to the NA61/SHINE experimental data~\cite{Porfy:2024jsu}. This opposite trend may be attributed to the onset of hydrodynamics in the experimental data, as the used UrQMD configuration lacks a hydrodynamic phase. Furthermore, the simulated results decrease monotonically, as opposed to the non-monotonic trend observed in the measured data~\cite{Porfy:2024jsu}.

\begin{figure}[]
\centerline{\includegraphics[width=1.0\textwidth]{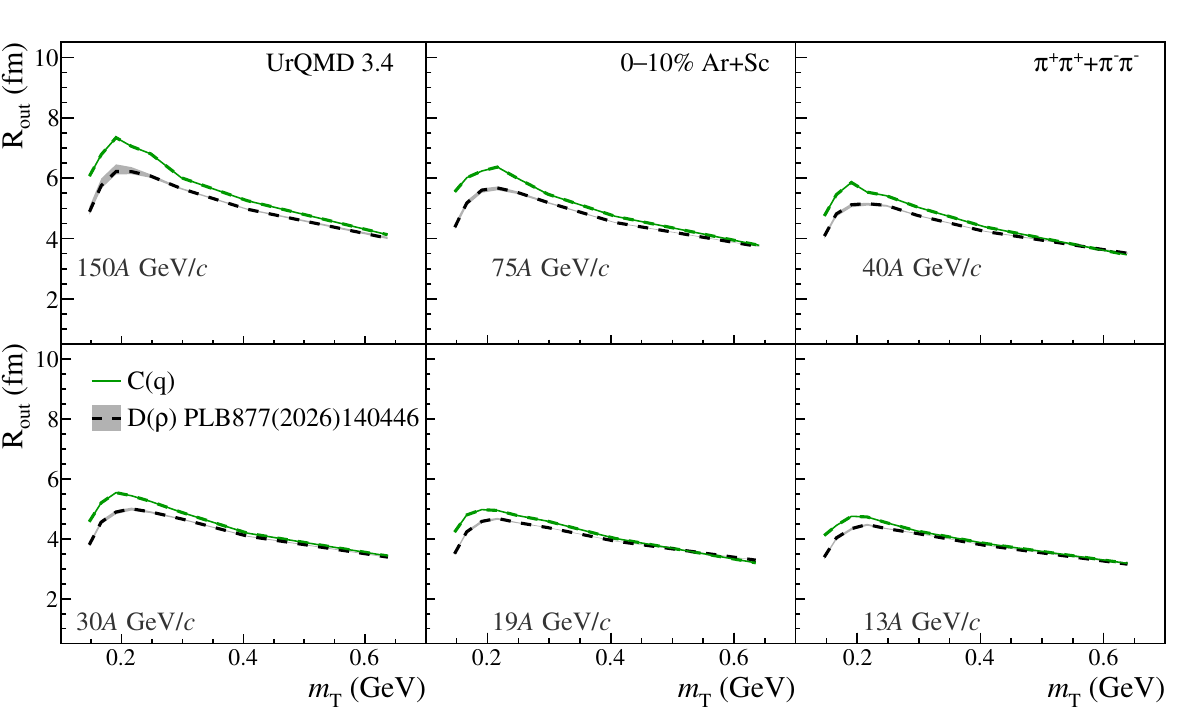}}
\caption{The Lévy scale parameter $R_\text{out}$ for 0--10\% central collisions in Ar+Sc as a function of $m_\text{T}$. The dashed lines represent values extracted from the simulated $D(\rho)$ distributions, while the solid lines represent the $C(q)$ results. The colored bands denote systematic uncertainties.}
\label{fig:fftvsDrhoRO}
\end{figure}

\begin{figure}[]
\centerline{\includegraphics[width=1.0\textwidth]{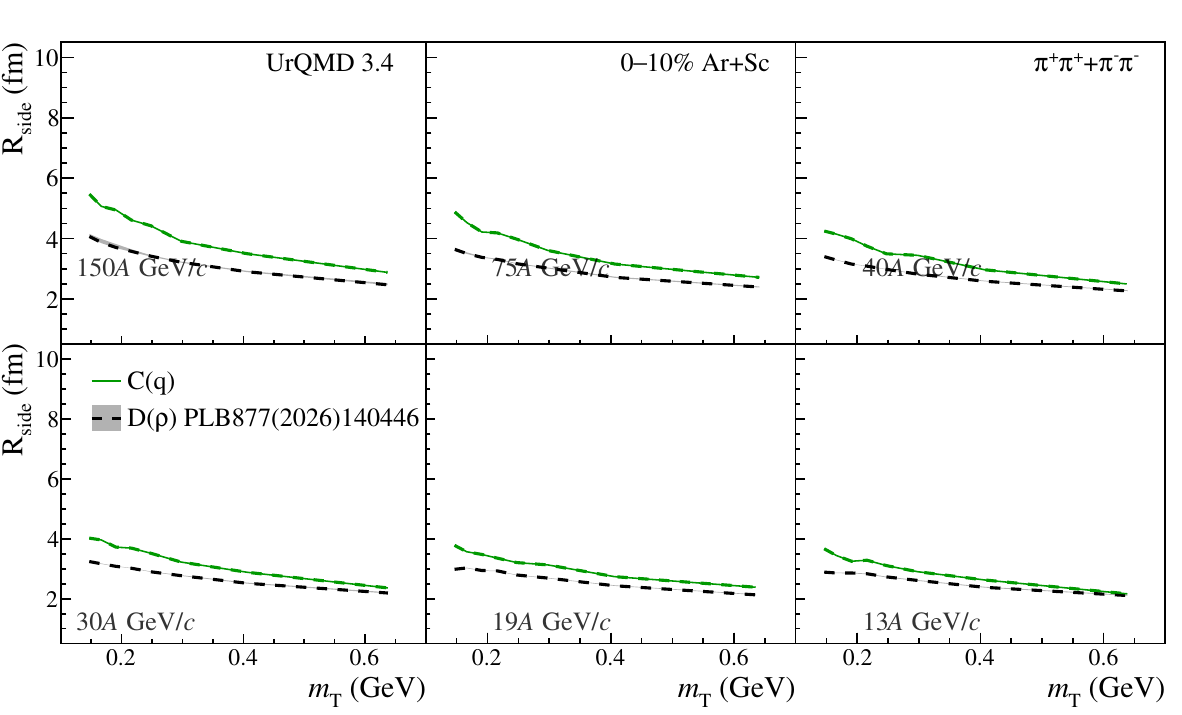}}
\caption{The Lévy scale parameter $R_\text{side}$ for 0--10\% central collisions in Ar+Sc as a function of $m_\text{T}$. The dashed lines represent values extracted from the simulated $D(\rho)$ distributions, while the solid lines represent the $C(q)$ results. The colored bands denote systematic uncertainties.}
\label{fig:fftvsDrhoRS}
\end{figure}

\begin{figure}[]
\centerline{\includegraphics[width=1.0\textwidth]{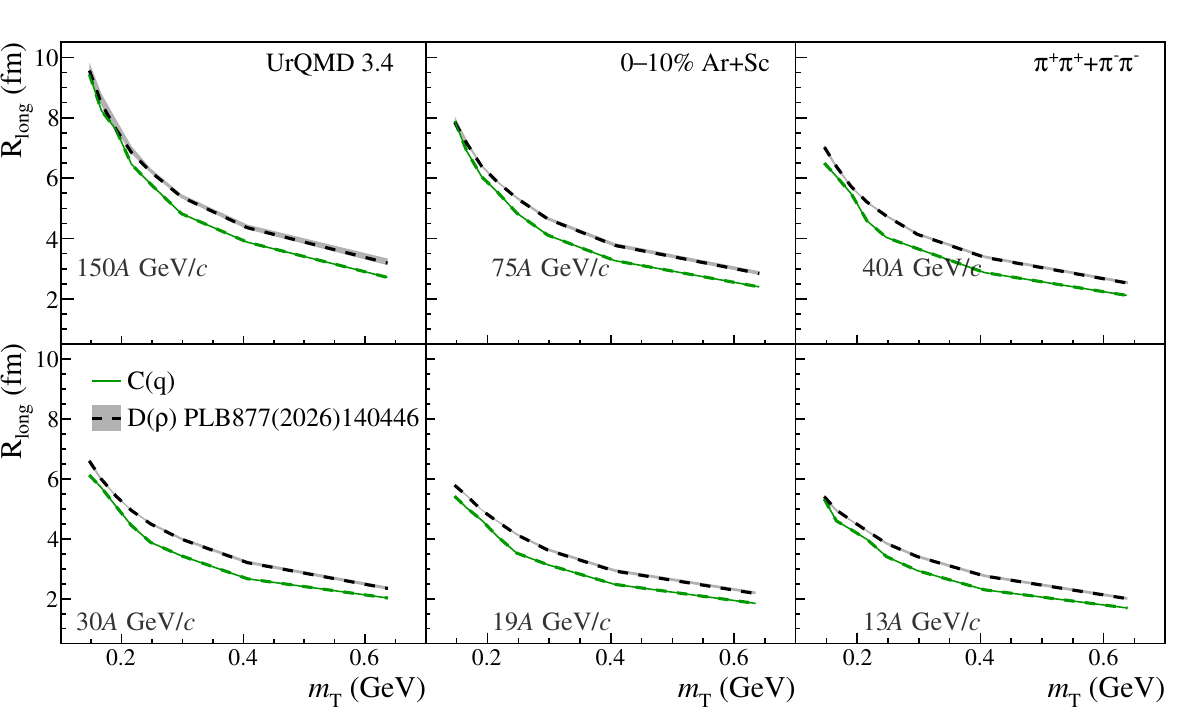}}
\caption{The Lévy scale parameter $R_\text{long}$ for 0--10\% central collisions in Ar+Sc as a function of $m_\text{T}$. The dashed lines represent values extracted from the simulated $D(\rho)$ distributions, while the solid lines represent the $C(q)$ results. The colored bands denote systematic uncertainties.}
\label{fig:fftvsDrhoRL}
\end{figure}

\begin{figure}[]
\centerline{\includegraphics[width=1.0\textwidth]{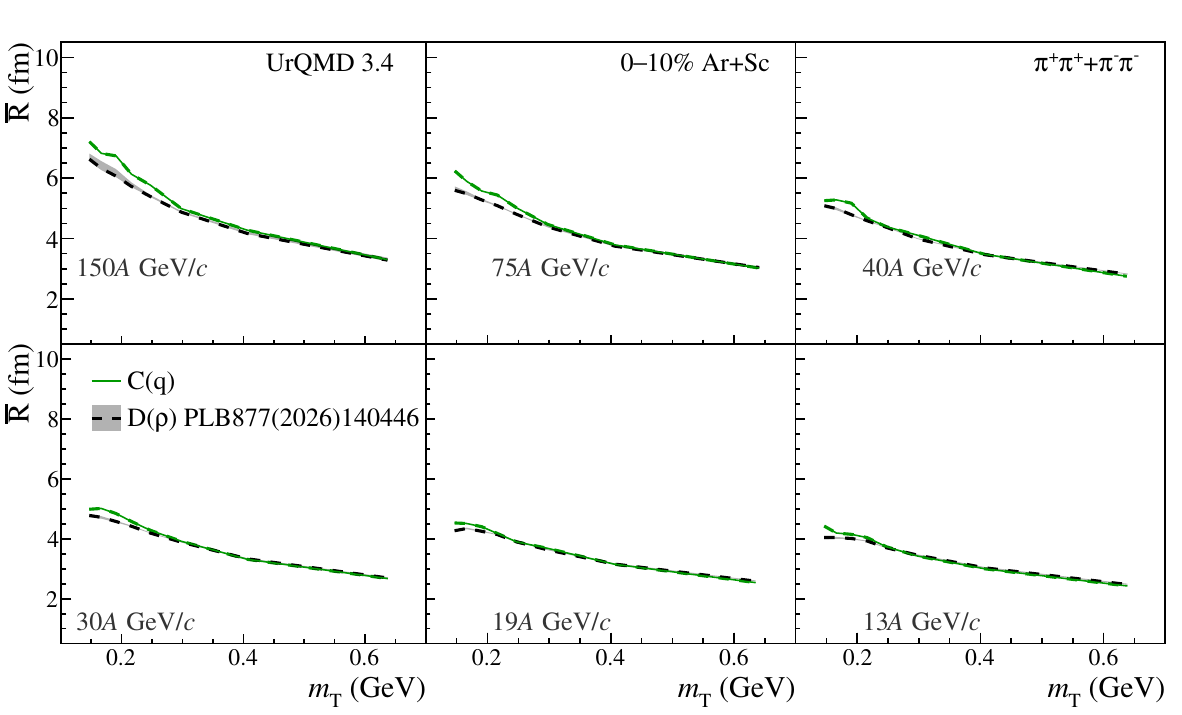}}
\caption{The Lévy scale parameter $\bar{R} = \sqrt{(R^2_\text{out} + R^2_\text{side} + R^2_\text{long}) / 3}$ for 0--10\% central collisions in Ar+Sc as a function of $m_\text{T}$. The dashed lines represent values extracted from the simulated $D(\rho)$ distributions, while the solid lines represent the $C(q)$ results. The colored bands denote systematic uncertainties.}
\label{fig:fftvsDrhoR}
\end{figure}

Moving to the Lévy scale parameters, $R^2_\text{out}$, $R^2_\text{side}$, and $R^2_\text{long}$, shown in Figs.~\ref{fig:fftvsDrhoRO}-\ref{fig:fftvsDrhoRL}, a clear decrease with $m_\text{T}$ can be observed, which can be attributed to an expanding source. In the ``out'' direction, two specific observations can be made: both methods feature a small maximum around 0.2 GeV/$c$, and the deviation between the methods decreases at lower collision energies. In a somewhat similar trend to $\alpha$, the $C(q)$ results for $R_\text{out}$ and $R_\text{side}$ converge systematically toward the $D(\rho)$ results at higher transverse masses, changing slightly over the energies. In the case of $R_\text{out}$ this convergence changes strongly with energy, specifically, at lower beam energies, the two results converge at lower transverse masses. Conversely, in the case of $R_\text{long}$, the parameter diverges from the $D(\rho)$ with increasing $m_\text{T}$. Finally, as a way of comparing three-dimensional results to one-dimensional experimental data of NA61/SHINE~\cite{Porfy:2024jsu}, the average of the three directions as $\bar{R} = \sqrt{(R^2_\text{out} + R^2_\text{side} + R^2_\text{long}) / 3}$ is also checked. As shown in Fig.~\ref{fig:fftvsDrhoR}, $\bar{R}$ exhibits a clear collision-energy dependence and demonstrates strong convergence between the two methods, indicating that the differences between the $D(\rho)$ and $C(q)$ approaches are less pronounced for this parameter.

\begin{figure}[]
\centerline{\includegraphics[width=1.0\textwidth]{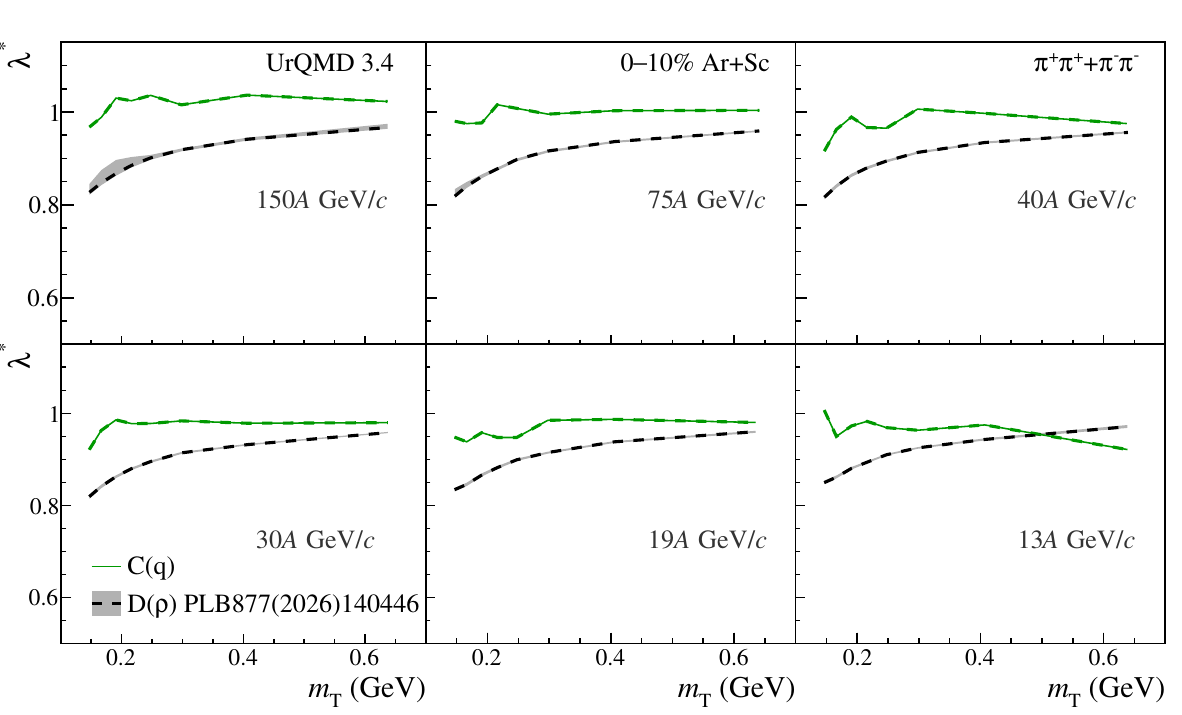}}
\caption{The intercept $\lambda^*$ for 0--10\% central collisions in Ar+Sc as a function of $m_\text{T}$. The dashed lines represent values extracted from the simulated $D(\rho)$ distributions, while the solid lines represent the $C(q)$ results. The colored bands denote systematic uncertainties.}
\label{fig:fftvsDrhoL}
\end{figure}

Finally, the intercept parameter is discussed, shown in Fig.~\ref{fig:fftvsDrhoL}. This parameter is denoted as $\lambda^*$ to differentiate it from the experimentally measured correlation strength $\lambda$, since the model lacks several long-lived weak decays (those of kaons, $\Sigma, \Omega, \Xi$ baryons), whose pions would contribute to the halo (hence $\lambda^*$ is close to or slightly larger than unity). A visible difference between the two methods can be seen: the parameter extracted from the correlation function shows a rapid increase in the first few $m_\text{T}$ bins, after which the values become approximately constant. The intercept parameter from $C(q)$ shows a quantitatively similar convergence as $\alpha$.

\section{Conclusion}
In this study, three-dimensional Lévy femtoscopic analyses were presented in central Ar+Sc collisions at SPS energies ($\sqrt{s_\text{NN}} \approx 5$--$17$ GeV) using UrQMD simulated events. The key goal of this work was to illustrate the difference between using the pair-source distribution or correlation function to estimate the Lévy fit parameters.

Parameters extracted directly from $D(\rho)$ fits are compared with those obtained from FFT-reconstructed correlation functions, $C(q)$, in the Ar+Sc system. Several differences can be observed, at low $m_\text{T}$, where $C(q)$ yields systematically smaller values for $\alpha$, $\lambda^*$, and $R_\text{long}$, whereas for $R_\text{out}$ and $R_\text{side}$ the opposite is seen. Interestingly, apart from $R_\text{long}$, the two methods converge strongly at higher $m_\text{T}$. In the case of the directionally averaged $\bar{R}$, the two methods provided approximately the same values.

Finally, comparing the simulated results to the experimental data highlights the limitations of the UrQMD model. The monotonic $m_\text{T}$ dependence of the Lévy index $\alpha$ observed in UrQMD is in direct contrast with the non-monotonic trend seen in experimental data. These results provide an important baseline for future measurements, however, they also highlight the need for the inclusion of the hydrodynamic phase for a more accurate description of the source.

\section*{Acknowledgments}
This work was supported by the NKFIH grant NKKP ADVANCED 152097 and NKFI grant 2025-1.1.5-NEMZ\_KI-2025-00003. Furthermore, the author was supported by DKÖP-23 Doctoral Excellence Program of the Ministry for Culture and Innovation and the EKÖP-26 University Research Scholarship Program of the Ministry of Science and Technology from the source of the National Research, Development and Innovation Fund.

\bibliographystyle{ws-ijmpa.bst}
\bibliography{wpcf}
\end{document}